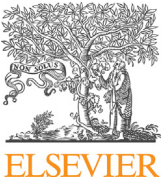
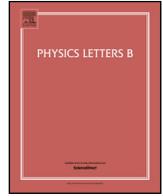

# Boltzmann factor and Hawking radiation

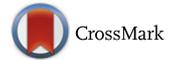

Gregory Ryskin

*Robert R. McCormick School of Engineering and Applied Science, Northwestern University, Evanston, IL 60208, United States*

**ABSTRACT**

Hawking radiation has thermal spectrum corresponding to the temperature $T_H = (8\pi M)^{-1}$, where $M$ is the mass (energy) of the black hole. Corrections to the Hawking radiation spectrum were discovered by Kraus and Wilczek (1995) and Parikh and Wilczek (2000). Here I show that these corrections follow directly from the basic principles of thermodynamics and statistical mechanics. In essence, it is the Boltzmann factor that ought to be corrected; corrections to the Hawking (or any other) radiation spectrum then follow necessarily.

© 2014 The Author. Published by Elsevier B.V. This is an open access article under the CC BY license (http://creativecommons.org/licenses/by/3.0/). Funded by SCOAP³.

Hawking radiation has thermal spectrum corresponding to the temperature $T_H = (8\pi M)^{-1}$, where $M$ is the mass (energy) of the black hole. (In Planck units, $G = c = \hbar = k = 1$.) This fact lies at the foundation of the black hole thermodynamics. Thus, when corrections to the Hawking radiation spectrum were discovered [1,2], they attracted a great deal of attention. Here I show that these corrections follow directly from the basic principles of thermodynamics and statistical mechanics. In essence, it is the Boltzmann factor that ought to be corrected; corrections to the Hawking (or any other) radiation spectrum then follow necessarily.

Consider a perfect black body in equilibrium with radiation. For the sake of simplicity, consider only those particles (quanta) in the radiation field whose chemical potential is zero, and whose energy $\epsilon \gg T$, where $T$ is the temperature of the body. For such particles, the probability $P_\alpha(1)$ that a single-particle quantum state $\alpha$ (the orbital $\alpha$) contains one particle is very small, $P_\alpha(1) \ll 1$, whereas the probability that it contains more than one particle is negligible for bosons, and zero for fermions. (The orbital is most likely to be empty; $P_\alpha(0) \approx 1$.) Then the mean occupation number $\bar{n}_\alpha$ of the orbital $\alpha$ is equal to $P_\alpha(1)$.

In equilibrium, the emission flux of particles of type $\alpha$ (such particles that can occupy the orbital $\alpha$) must equal their absorption flux; for a black body, the latter is proportional to the mean occupation number $\bar{n}_\alpha$. On the other hand, the emission flux is determined by the emission rate $\Gamma_\alpha$ of particles of type $\alpha$. ($\Gamma_\alpha$ is also called "the probability of emission per unit time", but this terminology is imprecise; it is $\Gamma_\alpha dt$ that has the meaning of probability. For sufficiently short intervals of time, the probability of emission of two or more particles of type $\alpha$ is negligible, whereas the probability of emission of one particle of type $\alpha$ is proportional to the duration of the interval, so $\Gamma_\alpha$ is well defined.) Thus in equilibrium the emission rate must satisfy

$$\Gamma_\alpha \sim P_\alpha(1), \qquad (1)$$

where the missing prefactor on the right-hand side has units of inverse time.

The ratio $P_\alpha(1)/P_\alpha(0)$ is given by the Boltzmann factor, $\exp(-\epsilon_\alpha/T)$. Since $P_\alpha(0)$ is very nearly equal to 1, this yields

$$P_\alpha(1) = \exp(-\epsilon_\alpha/T). \qquad (2)$$

Therefore

$$\Gamma \sim e^{-\epsilon/T}, \qquad (3)$$

where the subscript $\alpha$ has been dropped because this relation will apply to particles (quanta) of all types (with energy $\epsilon \gg T$ and zero chemical potential). The Hawking result fits this form, with temperature $(8\pi M)^{-1}$; this confirms the thermal nature of Hawking radiation.

Let us now recall the logic that leads to the Boltzmann factor. Consider a system $A$ in contact with another system ("the body"), together comprising an isolated system in thermodynamic equilibrium. The nature of the contact allows for free exchange of energy and particles. The basic principles of statistical mechanics are now invoked: the fundamental assumption is applied to the combined system (the body plus system $A$), while the Boltzmann postulate $S = \log \Omega$ is applied to the body (where $\Omega$ is the total number of accessible quantum states).

The fact – the event – of system $A$ being in the quantum state $j$ will be denoted $A \leftarrow j$. Its probability $P(A \leftarrow j)$ can be determined as follows: All the quantum states of the combined system are equally probable (the fundamental assumption). Therefore, the probability $P(A \leftarrow j)$ of system $A$ being in the quantum state $j$,

*E-mail address:* ryskin@northwestern.edu.

http://dx.doi.org/10.1016/j.physletb.2014.05.085
0370-2693/© 2014 The Author. Published by Elsevier B.V. This is an open access article under the CC BY license (http://creativecommons.org/licenses/by/3.0/). Funded by SCOAP³.

is proportional to the total number of those quantum states of the body that are compatible with $A \leftarrow j$ (i.e., with system $A$ being in the quantum state $j$). That is,

$$\frac{P(A \leftarrow i)}{P(A \leftarrow j)} = \frac{\exp(S_b|A \leftarrow i)}{\exp(S_b|A \leftarrow j)}, \quad (4)$$

where $S_b|A \leftarrow j$ is the entropy of the body given that system $A$ is in the quantum state $j$.

Eq. (4) can be rewritten as

$$\frac{P(A \leftarrow i)}{P(A \leftarrow j)} = \frac{\exp(\Delta S_b|A \leftarrow i)}{\exp(\Delta S_b|A \leftarrow j)}, \quad (5)$$

where $\Delta S_b|A \leftarrow j$ is equal to $S_b|A \leftarrow j$ minus the entropy of the body when system $A$ is in its "ground state" of zero energy and zero number of particles. That is, $\Delta S_b|A \leftarrow j$ is the change in the entropy of the body upon removing from it the number of particles and the amount of energy that system $A$ has in the quantum state $j$.

If only energy can be exchanged, the "ground state" of system $A$ will be that of zero energy, the number of particles in system $A$ being fixed. Then $\Delta S_b|A \leftarrow j$ is the change in the entropy of the body upon removing from it the amount of energy that system $A$ has in the quantum state $j$.

Eq. (5) can be rewritten in the form

$$P(A \leftarrow j) \propto e^{\Delta S_b|A \leftarrow j}. \quad (6)$$

It is clear from the derivation that Eq. (6) is valid generally, for any size of the body (and of system $A$); the only requirement being that the combined system can be viewed as isolated.

The rest of the analysis will be restricted to particles whose chemical potential is zero. (Relaxing this restriction leads to the Gibbs factor.) Then $\Delta S_b|A \leftarrow j$ does not depend on the number of particles even when particles can be exchanged. Since entropy is a monotonically increasing function of energy [3], $\Delta S_b|A \leftarrow j$ is a negative quantity.

If system $A$ is small compared to the body, its energy is likely to be small compared to the energy of the body. The entropy change of the body upon removal from it of a relatively small amount of energy $E$ can be expressed using the Taylor expansion, viz.,

$$\Delta S_b = -E \frac{\partial S_b}{\partial U_b} + \frac{E^2}{2} \frac{\partial^2 S_b}{\partial U_b^2} + \ldots = -\frac{E}{T} + \frac{E^2}{2} \frac{\partial}{\partial U_b} \frac{1}{T} + \ldots, \quad (7)$$

where $U_b$ is the energy of the body, and $T$ is its temperature. If the body is so large that it can be viewed as a heat reservoir, the second-order term and all the higher-order terms in Eq. (7) are zero by definition. Eq. (6) then implies the Boltzmann factor in the form

$$P(A \leftarrow j) \propto e^{-E_j/T}, \quad (8)$$

where $E_j$ is the energy of the quantum state $j$ of system $A$.

The above is essentially equivalent to the standard derivation of the Boltzmann factor (see, e.g., [4]). However, the intermediate step involving Eqs. (5), (6) is absent in the conventional derivation. In particular, Eq. (6) is of greater generality than the Boltzmann or Gibbs factor, yet has never been stated explicitly. (To the best of my knowledge. The exponential of $\Delta S$ in Eq. (6) resembles the Einstein formula for the probability of a fluctuation in the microcanonical ensemble [5], but the physical phenomena described by the two formulas are entirely different.)

Eq. (6) will now be used to derive a correction to the Boltzmann factor, as follows. If the body is not infinitely large, its temperature will change slightly upon transfer of energy to system $A$, and the Boltzmann factor must be corrected. This is easily accomplished by retaining the second-order term in Eq. (7). Let $C_b$ denote the heat capacity of the body. Eq. (7) can be rewritten as

$$\Delta S_b = -\frac{E}{T} - \frac{E^2}{2C_b T^2}, \quad (9)$$

leading, via Eq. (6), to the corrected Boltzmann factor in the form

$$\exp\left(-\frac{E}{T} - \frac{E^2}{2C_b T^2}\right). \quad (10)$$

Strictly speaking, $C_b < \infty$ for any physical system that can play the role of the body, so this result has universal applicability. Whether or not the correction is important will depend on the application.

In particular, such a correction can be calculated for a black hole in equilibrium with Hawking radiation, the black hole being the body, and the orbital $\alpha$ being system $A$. It is convenient to use the Hawking temperature $(8\pi M)^{-1}$ directly in Eq. (7); this yields (writing $\epsilon$ for $\epsilon_\alpha$)

$$\Delta S_{BH} = -8\pi M \epsilon + 4\pi \epsilon^2, \quad (11)$$

where $S_{BH} = 4\pi M^2$ is the Bekenstein–Hawking entropy of the black hole. Using now Eqs. (1), (6), and (11), one obtains the corrected emission rate

$$\Gamma \sim e^{\Delta S_{BH}} = e^{-8\pi \epsilon (M - \epsilon/2)}. \quad (12)$$

This should be compared to the results of the quantum theoretical calculations [1,2] in which corrections to the Hawking radiation spectrum were derived. (Ref. [1] has a factor of two error, corrected in [2].) Parikh and Wilczek [2] derived Hawking radiation as a tunneling process, and calculated the semiclassical emission rate. Their final result [2, Eq. (10)] is Eq. (12) in reverse order. That is, the relation $\Gamma \sim e^{\Delta S_{BH}}$ appears in Ref. [2] only *a posteriori*, as an alternative, succinct way to express the results (which can be lengthy – see [2, Eq. (16)]). The fundamental thermodynamic origin of this relation has not been understood hitherto; this is not surprising as Eq. (6) has not been available.

The corrections derived in [1,2] thus follow directly from the basic principles of thermodynamics and statistical mechanics. They are manifestations of a general result – it is the Boltzmann factor that ought to be corrected. The designation "deviations from thermality" [1,2,6] may not be the most appropriate.